\documentclass[twocolumn,showpacs,preprintnumbers]{revtex4}
\usepackage{graphicx}
\usepackage{dcolumn}
\usepackage{bm}
\usepackage{subfigure}
\begin{document}
\newcommand{\ds}{\displaystyle}
\newcommand{\be}{\begin{equation}}
\newcommand{\en}{\end{equation}}
\newcommand{\bea}{\begin{eqnarray}}
\newcommand{\ena}{\end{eqnarray}}
\title{Static spherically symmetric wormholes with isotropic pressure}
\author{Mauricio Cataldo}
\altaffiliation{mcataldo@ubiobio.cl} \affiliation{Departamento de
F\'\i sica, Facultad de Ciencias, Universidad del B\'\i o-B\'\i o,
Avenida Collao 1202, Casilla 15-C, Concepci\'on, Chile, and Grupo de
Cosmolog\'\i a y Gravitaci\'on-UBB.}
\author{Luis Liempi}
\altaffiliation{luiliempi@udec.cl} \affiliation{Departamento de
F\'\i sica, Universidad de Concepci\'on, Casilla 160-C,
Concepci\'on, Chile.\\}
\author{Pablo Rodr\'\i guez}
\altaffiliation{pablrodriguez@udec.cl} \affiliation{Departamento de
F\'\i sica, Universidad de Concepci\'on, Casilla 160-C,
Concepci\'on, Chile.\\}
\date{\today}
\begin{abstract}
In this paper we study static spherically symmetric wormhole
solutions sustained by matter sources with isotropic pressure. We
show that such spherical wormholes do not exist in the framework of
zero-tidal-force wormholes. On the other hand, it is shown that for
the often used power-law shape function there is no spherically
symmetric traversable wormholes sustained by sources with a linear
equation of state $p=\omega \rho$ for the isotropic pressure,
independently of the form of the redshift function $\phi(r)$. We
consider a solution obtained by Tolman at 1939 for describing static
spheres of isotropic fluids, and show that it also may describe
wormhole spacetimes with a power-law redshift function, which leads
to a polynomial shape function, generalizing a power-law shape
function, and inducing a solid angle deficit.

\vspace{0.5cm} 
\end{abstract}
\smallskip
\maketitle 

\section{Introduction}

In the framework of Einstein General Relativity the study of
spherically symmetric traversable wormhole
spacetimes~\cite{Morris:1988cz} has been mostly focused in matter
sources with anisotropic pressures. Mainly this is due to the fact
that in order to correctly describe a wormhole spacetime one needs a
redshift function without horizons, or the redshift and the shape
functions giving a desired asymptotic. In this way, the theoretical
construction of wormhole geometries is usually performed by assuming
a priori the form of the redshift and the shape functions, in order
to have a desired metric. But Einstein's field equations for
spherically symmetric spacetimes imply that the radial and lateral
(or transverse) pressures are not equal. In such a way, by imposing
restricted choices on the redshift and the shape functions we will
obtain, in general, for the energy-momentum tensor that $T_2^2 \neq
T_3^3$. This condition implies that a wormhole configuration is
necessarily supported by an anisotropic matter source. However, this
method has some limitations since we can obtain for the energy
density, radial and lateral pressures algebraic expressions which
are physically unreasonable.

One may also follow a more conventional method used for finding
solutions in general relativity, by prescribing the matter content
with specific equations of state for the radial and/or the
tangential pressures, and then solve Einstein's field equations in
order to find the redshift and shape functions. The often used
equation of state is the linear barotropic equation of state
$p_r=\omega \rho$, which relates the radial pressure with the energy
density~\cite{Kuhfittig}. From the cosmological setting, such an
equation of state is associated with phantom dark energy if
$\omega<-1$, and also may sustain traversable static
wormholes~\cite{Sushkov:2005kj,Lobo}.

Anisotropic stress contributions to the gravitational field can
arise from specific matter fields. A fluid source with anisotropic
stresses supporting wormholes may be for example of electromagnetic
nature: linear~\cite{Balakin} and nonlinear Maxwell
fields~\cite{Hendi} have been considered in the literature. An
electric field coupled to a scalar field is considered in
Ref.~\cite{Kim}.

A spatially varying cosmological constant also has been considered
in the framework of static wormholes supported by anisotropic matter
sources~\cite{Rahaman}. In Ref.~\cite{DeBenedictis} the source of
the stress-energy tensor supporting the wormhole geometries consists
of an anisotropic brown dwarf "star" which smoothly joins the vacuum
and may possess an arbitrary cosmological constant, while in
Ref.~\cite{Popov} anisotropic vacuum stress-energy of quantized
fields has been proposed as source for static wormholes.

Such anisotropic scenarios are obtained also for regular static,
spherically symmetric solutions describing wormholes supported by
dark matter non-minimally coupled to dark energy in the form of a
quintessence scalar field~\cite{Folomeev}.

The main purpose of this paper is to present and discuss static
spherically symmetric wormhole spacetimes supported by a single
perfect fluid, i.e. a matter source with isotropic pressure. As far
as we know, the only spherical wormhole solution discussed up to now
is the not asymptotically flat wormhole with isotropic pressure
considered in Ref.~\cite{Lobo}. We will discuss this solution in
more detail below in Sections IV and V. It must be noticed that the
study of wormhole solutions sustained by a perfect fluid allows us
to consider phantom wormholes sustained by inhomogeneous and
isotropic phantom dark energy.

The paper is organized as follows. In Sec. II we write the Einstein
equations for static spherically symmetric spacetimes. In Sec. III
we analyze the possibility of having zero-tidal-force wormholes
sustained by a matter source with isotropic pressure, while in Sec.
IV we analyze the possibility of having spherical wormholes
sustained by a fluid with linear equation of state. In Sec. V we
re-obtain an analytical solution, previously obtained by Tolman,
which describes static spheres of fluids with isotropic pressure,
and we show that it may describe also a non-asymptotically flat
wormhole geometry.

\section{Field equations for static spherically symmetric spacetimes}
The spacetime ansatz for seeking static spherically symmetric
solutions can be written in Schwarzschild coordinates as
\begin{eqnarray}\label{general BH metric}
ds^2=e^{2 \phi(r)} dt^2- \frac{dr^2}{1-\frac{b(r)}{r}}-r^2 \left(
d\theta^2+\sin^2\theta d\phi^2 \right),
\end{eqnarray}
where $e^{\phi(r)}$ and $b(r)$ are arbitrary functions of the radial
coordinate. In the case of wormholes these functions are referred to
as redshift function and shape function respectively. The essential
characteristics of a wormhole geometry are encoded in these
functions, so in order to have a wormhole these two functions must
satisfy some general constraints discussed by Morris and Thorne in
Ref.~\cite{Morris:1988cz,Visser}.

By assuming that the matter content is described by a single
imperfect fluid, from the metric~(\ref{general BH metric}) and the
Einstein field equations $G_{\mu \nu}=-\kappa T_{\mu \nu}$ we obtain
\begin{eqnarray}
\kappa \rho(r)=\frac{b^{\prime}}{r^2}, \label{rho} \\
\kappa p_r(r)=2\left(1-\frac{b}{r}  \right)
\frac{\phi^{\prime}}{r}-\frac{b}{r^3}, \label{pr} \\
\kappa p_l(r)=\left( 1-\frac{b}{r} \right) \times \nonumber
\\ \left[\phi^{\prime \prime}+\phi^{\prime \, 2} - \frac{b^\prime
r+b-2r}{2r(r-b)} \, \phi^\prime - \frac{b^\prime
r-b}{2r^2(r-b)}\right], \label{pl}
\end{eqnarray}
where $\kappa=8 \pi G$, $\rho$ is the energy density, and $p_r$ and
$p_l$ are the radial and lateral pressures respectively. From the
conservation equation $T^{\mu \nu}_{\,\,\,\,\,\,\, ; \nu}=0$ we
obtain the hydrostatic equation for equilibrium of the matter
sustaining the wormhole
\begin{eqnarray}\label{conservation equation}
p_r^\prime=\frac{2(p_l-p_r)}{r}- (\rho+p_r) \phi^\prime.
\end{eqnarray}

It becomes clear that the main condition for having a perfect fluid
is given by
\begin{equation}\label{IC15}
p_r=p_l.
\end{equation}
This condition on the radial and lateral pressures allows us to get
the following differential equation connecting functions $\phi(r)$
and $b(r)$:
\begin{eqnarray}\label{isotropic condition}
\phi^{\prime \prime}+\phi^{\prime \, 2} - \frac{b^\prime
r-3b+2r}{2r(r-b)} \, \phi^\prime = \frac{b^\prime r-3b}{2r^2(r-b)}.
\end{eqnarray}
We may consider Eq.~(\ref{isotropic condition}) as a differential
equation for one of these involved functions, by giving the
remaining one. By supposing that the redshift function $\phi(r)$ is
given, we obtain a first order differential equation for the shape
function $b(r)$, whose general solution is given by
\begin{widetext}
\begin{eqnarray}\label{isotropic integral}
b(r)=\left ( \int\frac{2r \left( r\phi^{\prime \prime}+r\phi^{\prime
2}-\phi^\prime\right) e^{\int \frac{2r^2\phi^{\prime
\prime}+2r^2\phi^{\prime 2}-3r\phi^\prime-3}{r(1+r
\phi^\prime)}dr}}{1+r \phi^\prime}\, dr +C\right) \, e^{-\int
\frac{2r^2\phi^{\prime \prime}+2r^2\phi^{\prime
2}-3r\phi^\prime-3}{r(1+r \phi^\prime)} dr},
\end{eqnarray}
\end{widetext}
where $C$ is an integration constant. Eqs.~(\ref{isotropic
condition}) and~(\ref{isotropic integral}) have a general character,
in the sense that they do not involve an equation of state for
$\rho$ and $p$. It must be remarked that for these static
configurations, sustained by isotropic perfect fluids, the Einstein
field equations are reduced to a set of three independent
differential equations~(\ref{rho}), (\ref{pr}) and~(\ref{isotropic
condition}) for four unknown functions, namely $\phi(r)$, $b(r)$,
$\rho(r)$ and $p(r)$. Thus, to study solutions to these field
equations, restricted choices of one of the unknown functions must
be considered.

To obtain a realistic stellar model, one can start with an equation
of state. Such input equations of state do not normally allow for
closed form solutions. In arriving to exact solutions, one can solve
the field equations by making an ad hoc assumption for one of the
metric functions or for the energy density. Hence the equation of
state can be computed from the resulting metric.

\section{On zero-tidal-force wormholes with isotropic pressure}

It is well known that a simple class of solutions corresponds to
zero-tidal-force wormhole spacetimes, which are defined by the
condition $\phi(r)=\phi_0=const$~\cite{Morris:1988cz,Visser}. By
putting $\phi(r)=const$ into Eq.~(\ref{isotropic integral}) we
obtain $b(r)=Cr^3$. By requiring that $b(r_0)=r_0$, the spacetime
metric takes the form
\begin{eqnarray}\label{CC solution}
ds^2= dt^2- \frac{dr^2}{1-\left(\frac{r}{r_0}\right)^2}-r^2 \left(
d\theta^2+\sin^2\theta d\phi^2 \right).
\end{eqnarray}
This metric represents a spacetime of constant curvature, for which
the pressure and energy density are given by $p=-\rho/3=-1/\kappa
r_0^2$, and it is a particular case of the well-known static
Einstein universe (for which we have $\kappa \rho=3/r_0^2 -\Lambda$
and $\kappa p =-1/r_0^2+\Lambda$, where $\Lambda$ is the
cosmological constant).

The inverse of the radial metric component $g_{rr}^{-1}$ vanishes at
$r=r_0$, as we would expect for wormholes. However, for $r>r_0$ the
radial metric component $g_{rr}$ becomes negative, so the solution
is valid only for $0\leq r \leq r_0$. This implies that there is no
zero-tidal-force wormhole solutions sustained everywhere by an
isotropic perfect fluid. In other words, any zero-tidal-force
wormhole must be filled by a single fluid with anisotropic pressure.
Therefore, in order to generate spherically symmetric wormholes,
sustained by a single matter source with isotropic pressure, we must
consider spacetimes with $\phi(r) \neq const$.

Notice that this result does not mean that it is not possible to
have a zero-tidal-force wormhole sustained by a perfect (ideal)
fluid at spatial infinity. For an explicit example let us consider
the spacetime
\begin{eqnarray}\label{dos}
ds^2=dt^2-\frac{dr^2}{\left(\frac{r}{r_0}\right)^\alpha-1}-r^2
\left( d\theta^2+\sin^2\theta d\phi^2 \right).
\end{eqnarray}
For $\alpha>0$ the metric covers the range $r_0 \leq r < \infty$ and
describes a wormhole spacetime. The energy density, radial and
lateral pressures are given by $\kappa
\rho=-\frac{1+\alpha}{r_0^{2}}\left(\frac{r}{r_0}
\right)^{\alpha-2}+\frac{2}{r^2}$, $\kappa
p_r=\frac{1}{r_0^{2}}\left(\frac{r}{r_0}
\right)^{\alpha-2}-\frac{2}{r^2}$ and $\kappa
p_l=\frac{\alpha}{2r_0^{2}}\left(\frac{r}{r_0} \right)^{\alpha-2}$.
Since $\phi(r)=0$ this wormhole has anisotropic pressures as we
would expect. The metric~(\ref{dos}) includes non-asymptotically
flat spacetimes. Note that for $\alpha=2$ we have that
$\rho=-3/r_0^2+2/r^2$, $p_r=1/r_0^2-2/r^2$ and $p_l=1/r_0^2$,
obtaining at spatial infinity a spacetime with constant curvature
and $p_r=p_l=-\rho/3$, which means that asymptotically we have a
wormhole sustained by an ideal fluid (a string gas).



\section{On spherical wormholes with isotropic pressure and linear
equation of state}

Since, for a vanishing redshift function, the only solution allowed
by the isotropic pressure condition~(\ref{IC15}) is the
spacetime~(\ref{CC solution}), we shall now on take into
consideration a non-vanishing redshift function in all considered
wormhole solutions.

In this section, in order to follow with our study, we shall impose
on the isotropic pressure the linear equation of state
\begin{eqnarray} \label{barotropic equation}
p=\omega \rho,
\end{eqnarray}
where the state parameter $\omega$ is a constant. Thus from
Eq.~(\ref{conservation equation}) we have
\begin{eqnarray}\label{conservation equation para pomegarho}
\omega \rho^\prime=-(1+\omega)\rho \phi^\prime
\end{eqnarray}
and then the energy density is given by
\begin{eqnarray*}\label{rhointegrada}
\rho(r)=C e^{-\frac{1+\omega}{\omega}\phi(r)},
\end{eqnarray*}
where $C$ is an integration constant. If $\phi(r)=const$ we obtain
that energy density is constant, in agreement with the result of the
previous section.

From Eqs.~(\ref{rho}), (\ref{isotropic condition})
and~(\ref{conservation equation para pomegarho}) the following
master differential equation for the shape function $b(r)$ is
obtained:
\begin{widetext}
\begin{eqnarray}\label{ecuacion para b}
-2r^2 \omega (1+\omega)(r-b)b^\prime b^{\prime \prime \prime} +4
\omega \left(\omega +\frac{1}{2} \right) r^2 (r-b) b^{\prime \prime
2}+\omega r ((1+\omega) r b^\prime+ (5 \omega -3) b+2r-6\omega r)
b^\prime b^{\prime \prime}- \nonumber \\ 3 b^{\prime 2}
\left(\left(\omega+\frac{1}{3}\right)(1+\omega)r b^\prime -
\left(1+\frac{5}{3} \omega^2+\frac{16}{3} \omega
\right)b+\frac{8}{3} \omega r \right)=0.
\end{eqnarray}
\end{widetext}
In general it is hard to find analytical solutions to
Eq.~(\ref{ecuacion para b}). Nevertheless, one can make some checks
to prove the correctness of the above equation. Notice, for example,
that Eq.~(\ref{ecuacion para b}) is fulfilled identically for
$b(r)=Ar^3$. Thus from Eq.~(\ref{conservation equation para
pomegarho}) we obtain that $\omega=-1$ or $\phi(r)=const$. For the
latter case $p=-\rho/3=-A$, while for $\omega=-1$ we have that
$e^{\phi(r)}=1-A r^2$ with $p=-\rho=-3A$.

On the other hand, the first term of Eq.~(\ref{ecuacion para b})
vanishes for $\omega=0$ and $\omega=-1$. Thus for $\omega=0$ we
obtain $b(r)=A$, $e^{2 \phi(r)}=1-A/r$, $\rho=p=0$, i.e. the
Schwarzschild solution. It is well known that the Schwarzschild
solution may be interpreted as a non-traversable wormhole.

For $\omega=-1$ we obtain the Kottler solution, i.e. $b(r)=A+B r^3$,
$e^{2 \phi(r)}=1-A/r-B r^2$, $\rho=-p=3B$~\cite{Kottler}.

We turn next to often used power-law form of the shape function in
wormhole spacetimes: $b(r)=A/r^n$. This choice ensures that for $r
\rightarrow \infty$ and $n>-1$ the M-T constraint $b(r)/r \leq 1$ is
satisfied. Thus, by putting $b(r)=A/r^n$ into the master
equation~(\ref{ecuacion para b}) we find that this equation is
satisfied if $n=-3$ (for arbitrary $\omega$) or $n=-5/3$ (and
$\omega=-3$).

The obtained negative values of the $n$-parameter are less than
$-1$. This implies that there is no spherically symmetric
traversable wormholes characterized by a radial metric component
given by $g_{rr}^{-1}=1-(r_0/r)^{n+1}$ (with $n>-1$), and sustained
by isotropic pressure sources with a linear equation of
state~(\ref{barotropic equation}), independently of the form of the
redshift function $\phi(r)$. The implications of this result tell us
that we must consider more general forms of the shape function
$b(r)$ and/or of the equation of state of the isotropic pressure
$p(\rho)$.

Lastly, let us note that this conclusion is not in agreement with
the result obtained in subsection III-B of the Ref.~\cite{Lobo},
where the author discusses the non-asymptotically flat wormhole
given by
\begin{eqnarray}
ds^2=(r/r_0)^{2\omega \left(
\frac{3-\alpha}{1+\omega}\right)}dt^2-\frac{dr^2}{1-(r_0/r)^{1-\alpha}}-
\nonumber \\ r^2 \left( d\theta^2+\sin^2\theta d\phi^2 \right), \\
p=\omega \rho=-\frac{1}{8 \pi r_0^2} \left(\frac{r_0}{r}
\right)^{3-\alpha}, \label{LES}
\end{eqnarray}
where $\alpha=-1/\omega$ (see Eqs. (32) and (33) of
Ref.~\cite{Lobo}). Notice that the equation of state~(\ref{LES}) has
the linear form of Eq.~(\ref{barotropic equation}) and the shape
function just the form not allowed by the master Eq.~(\ref{ecuacion
para b}), therefore this non-asymptotically flat solution is not
consistent with Einstein equations~(\ref{rho})-(\ref{pl}). It
becomes clear that this solution is defined by a redshift function
of general form given by
\begin{eqnarray}\label{ephi}
e^{\phi(r)}=\left(\frac{r}{r_0} \right)^\beta,
\end{eqnarray}
with $\beta$ a constant. In the following section we shall discuss
the solution generated by field equations~(\ref{rho})-(\ref{pl})
with the restricted choice~(\ref{ephi}).

\section{On spherical wormhole with
$e^{\phi(r)}=\left(\frac{r}{r_0} \right)^{\beta}$}

In order to show the correctness of the conclusion of the previous
section, we must provide the static spherically symmetric solution
with the specific choice of the power-law redshift
function~(\ref{ephi}). In Ref.~\cite{Tolman} Tolman provides
explicit analytical solutions for static spheres of fluids with
isotropic pressure. For our purpose, it is convenient to consider
the solution V, obtained in Sec. 4 of Ref.~\cite{Tolman}, which
justly take into account the metric component $g_{tt}$ having the
form of Eq.~(\ref{ephi}). We shall re-obtain the Tolman solution by
assuming that the redshift function is given by Eq.~(\ref{ephi}).

By putting Eq.~(\ref{ephi}) into Eq.~(\ref{isotropic condition}) we
find for the shape function
\begin{eqnarray}\label{bder}
b(r)={\frac {\beta\, \left( \beta-2 \right) }{\beta^2-2\,\beta-1}}
\, r  - C \, {r}^{-{\frac { \left( 2\,\beta+1 \right) \left( -
3+\beta \right) }{1+\beta}}},
\end{eqnarray}
where $C$ is a constant of integration. It becomes clear that this
form of the shape function is more general than the discussed above
power-law shape function $b(r)=A/r^n$, as we would expect. Thus, the
metric, energy density and pressure are provided by
\begin{widetext}
\begin{eqnarray}\label{metric}
ds^2&=&\left(\frac{r}{r_0} \right)^{2 \beta} dt^2-
\frac{dr^2}{1-{\frac {\beta\, \left( \beta-2 \right)
}{\beta^2-2\,\beta-1}}+\tilde{C} \, {\left(\frac{r}{r_0}
\right)}^{-{\frac {2(\beta^2-2 \beta-1)}{1+\beta}}}}-
r^2 \left( d\theta^2+\sin^2\theta d\phi^2 \right),  \\
\kappa \rho(r)&=& \tilde{C} \, \left( {\frac {r}{r_{{0}}}} \right)
^{-2\,{\frac {-2\, \beta-1+{\beta}^{2}}{1+\beta}}} \left( 2\,\beta+1
\right) \left( \beta-3 \right)  \left( 1+\beta \right)
^{-1}{r}^{-2}+{\frac { \left( \beta-2 \right) \beta}{ \left(
-2\,\beta-1+{\beta}^{2} \right) {r}^{2} }}, \\
\kappa p(r)&=& \left( 2\,\beta+1 \right) {\tilde{C}}\, \left( {\frac
{r}{r_{{0}}}}
 \right) ^{-2\,{\frac {-2\,\beta-1+{\beta}^{2}}{1+\beta}}}{r}^{-2}-{
\frac {{\beta}^{2}}{ \left( -2\,\beta-1+{\beta}^{2} \right)
{r}^{2}}},
\end{eqnarray}
\end{widetext}
respectively, where $\tilde{C}=C
r_0^{{-2(\beta^2-2\beta-1)}/(1+\beta)}$. Note that the pressure is
isotropic but not of barotropic type. We have such an equation of
state only for $\tilde{C}=0$ and $\beta=-1/2$, obtaining for the
latter value $g_{tt}=r_0/r$, $g_{rr}^{-1}=\tilde{C}r_0/r-4$, and
$-p=\rho/5=\frac{1}{ \kappa r^2}$, i.e. $\omega=-1/5$. This solution
exhibits a solid angle deficit, and does not describe a wormhole
geometry.

By requiring the standard wormhole condition $b(r_0)=r_0$ on
Eq.~(\ref{bder}) we obtain for the metric~(\ref{metric}) the
following expression:
\begin{eqnarray}\label{metric SWH MT}
ds^2=\left(\frac{r}{r_0} \right)^{2 \beta} dt^2- \frac{(1+2 \beta
-\beta^2) \, dr^2}{1-{\left(\frac{r}{r_0} \right)}^{{\frac {2(1+2
\beta -\beta^2)}{1+\beta}}}}- \nonumber
\\ r^2 \left( d\theta^2+\sin^2\theta d\phi^2 \right).
\end{eqnarray}


In order to the metric~(\ref{metric SWH MT}) describes a traversable
wormhole the condition ${\frac{1+2\beta-\beta^2}{1+\beta}}<0$ must
be required, implying that the parameter $\beta$ varies in the
ranges $\beta>1+\sqrt{2}$ or $-1<\beta<1-\sqrt{2}$. However, in this
case $1+2\beta-\beta^2<0$, so the radial metric component $g_{rr}$
becomes negative, and then the metric~(\ref{metric SWH MT}) does not
describes a wormhole solution.

The metric component $g_{rr}$ does not change its sign if we require
${\frac{1+2\beta-\beta^2}{1+\beta}}>0$ and $1+2\beta-\beta^2>0$.
This implies that the parameter $\beta$ varies in the ranges
$1-\sqrt{2} <\beta<1+\sqrt{2}$. In this case the radial coordinate
varies from zero to a maximum value $r_{max}=r_0>0$, hence the
solution corresponds to a fluid sphere of radius $r_0$ with
isotropic pressure.

Nevertheless, as we shall see in the following subsection, the
considered Tolman solutions can describe wormhole geometries
fulfilling the required conditions $g_{rr}>0$ and $r \geq r_0$.

\subsection{A truly wormhole geometry}

In order to show that the metric~(\ref{metric SWH MT}) may correctly
describe a Lorentzian spacetime for $r \geq r_0$ let us rewrite it
in the following form:
\begin{eqnarray}\label{metric LM MT}
ds^2=\left(\frac{r}{r_0} \right)^{2 \beta} dt^2-
\frac{(\beta^2-2\,\beta-1) \, dr^2}{{\left(\frac{r}{r_0}
\right)}^{{\frac {2(1+2 \beta -\beta^2)}{1+\beta}}}-1}- \nonumber
\\ r^2 \left( d\theta^2+\sin^2\theta d\phi^2 \right).
\end{eqnarray}
It becomes clear that this metric describes a Lorentzian spacetime
for $r \geq r_0$ if $\beta^2-2\,\beta-1>0$ and $\frac {1+2 \beta
-\beta^2}{1+\beta}>0$, which implies that the condition $\beta<-1$
must be required.

For the metric~(\ref{metric LM MT}) the energy density and the
isotropic pressure are given by
\begin{eqnarray}\label{ED}
\kappa \rho=\frac{\left(\frac{r}{r_0}
\right)^{-\frac{2(\beta^2-2\beta-1)}{1+\beta}}(2\beta+1)(\beta-3)}{(1+\beta)(\beta^2-2\beta-1)r^2}+
\\ \nonumber \frac{\beta(\beta-2)}{(\beta^2-2\beta-1)r^2}, \\ \label{PL}
\kappa p=\frac{\left(\frac{r}{r_0}
\right)^{-\frac{2(\beta^2-2\beta-1)}{1+\beta}}(2\beta+1)}{(\beta^2-2\beta-1)r^2}-\\
\nonumber \frac{\beta^2}{(\beta^2-2\beta-1)r^2}.
\end{eqnarray}
It is interesting to note that this geometry describes a spacetime
with a solid angle deficit (or excess). This can be seen directly by
making the rescaling $\varrho^2=(\beta^2-2\beta-1)r^2$. Then, the
metric~(\ref{metric LM MT}) becomes
\begin{eqnarray}\label{metric LM MT varrho}
ds^2=\left(\frac{\varrho}{\varrho_0} \right)^{2 \beta} dt^2- \frac{
d\varrho^2}{{\left(\frac{\varrho}{\varrho_0} \right)}^{{\frac {2(1+2
\beta -\beta^2)}{1+\beta}}}-1}- \nonumber
\\ \frac{\varrho^2}{(\beta^2-2\,\beta-1)} \left( d\theta^2+\sin^2\theta d\phi^2 \right).
\end{eqnarray}
This new form of the metric~(\ref{metric LM MT}) shows explicitly
the presence of a solid angle deficit for $-\infty<\beta<1-\sqrt{3}$
or $1+\sqrt{3}<\beta<\infty$, and a solid angle excess for
$1-\sqrt{3}<\beta<1-\sqrt{2}$ or $1+\sqrt{2}<\beta<1+\sqrt{3}$.
These topological defects vanish for $\beta=1 \pm \sqrt{3}$,
obtaining a non-flat asymptotic spacetime.

\begin{figure}
\includegraphics[scale=0.38]{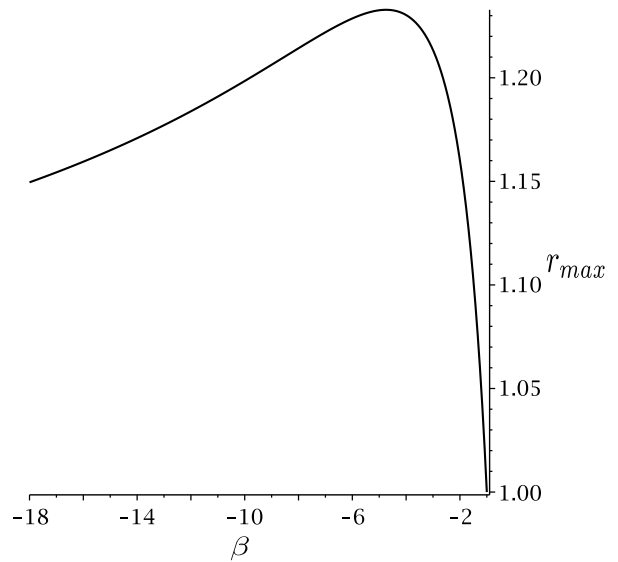}
\caption{Plot shows the behaviour of $r_{max}$ of Eq.~(\ref{rmax})
for $\beta \leq -1$ and $r_0=1$. The maximum value is reached at
$\beta=-4.745695219$ where $r_{max}=1.232835973$. For $\beta
\rightarrow -\infty$ we have $r_{max} \rightarrow 1$.}
\label{Graficoparabeta15}
\end{figure}

\begin{figure}
\includegraphics[scale=0.389]{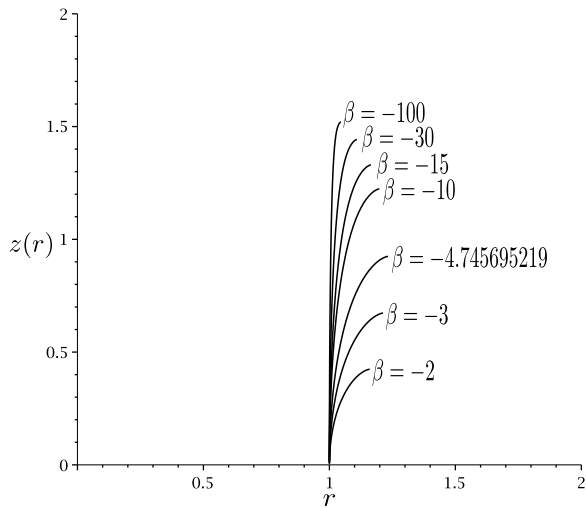}
\caption{Plots of the embedding function $z(r)$ for various values
of the $\beta$-parameter are shown. The throat width has been set to
$r_0=1$. The embedding of each slice $t=const,
\theta=\frac{\pi}{2}$, $\beta=const$ of the wormhole~(\ref{metric LM
MT}) extends from the throat $r_0=1$ to $r_{max}>r_0$. The heigh of
the $z$-function increases with decreasing $\beta$-parameter. It
becomes clear that as $\beta \rightarrow -\infty$ the $r_{max}
\rightarrow r_0=1$.} \label{embedding-lineas15}
\end{figure}

\begin{figure}
\includegraphics[scale=0.1415]{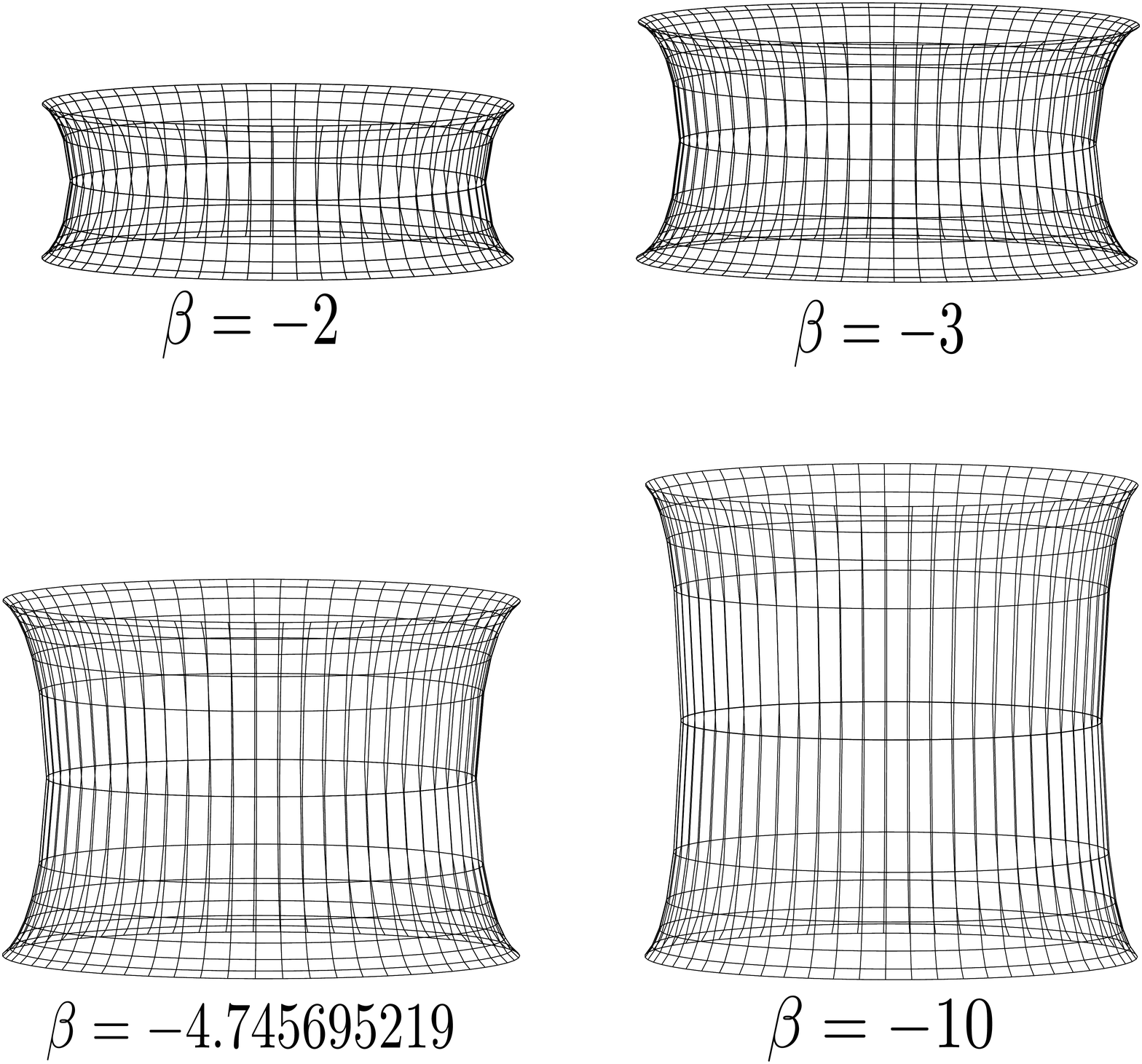}
\caption{The figure shows three dimensional wormhole embedding
diagrams for $\beta=-2,-3,-4.745695219,-10$. The heighs of the
diagrams increase with decreasing $\beta$-parameter. The throat
width is the same for all diagrams.} \label{beta-615}
\end{figure}

The geometrical properties and characteristics of these solutions
can be explored through the embedding diagrams, which helps to
visualize the shape and the size of slices $t=const,
\theta=\frac{\pi}{2}$ of the metric~(\ref{metric LM MT}) by using a
standard embedding procedure in ordinary three dimensional Euclidean
space. In general, in order to embed two dimensional slices
$t=const, \theta=\frac{\pi}{2}$ of the generic metric~(\ref{general
BH metric}) the equation
\begin{eqnarray}
\frac{dz(r)}{dr}=\frac{1}{\sqrt{\frac{r}{b(r)}-1}}
\end{eqnarray}
is used for the lift function $z(r)$~\cite{Morris:1988cz}. Thus, for
slices $t=const, \theta=\frac{\pi}{2}$ of the metric~(\ref{metric LM
MT}) we obtain for the first derivative of the lift function

\begin{eqnarray}\label{embedding}
\frac{dz(r)}{dr}=\sqrt{\frac{\beta(\beta-2)-\left(\frac{r}{r_0}
\right)^{\frac{2(1+2\beta-\beta^2)}{\beta+1}}}{\left(\frac{r}{r_0}
\right)^{\frac{2(1+2\beta-\beta^2)}{\beta+1}}-1}}.
\end{eqnarray}

This expression implies that at the throat
$\frac{dz(r)}{dr}=\infty$, and it vanishes for $\beta=1 \pm
\sqrt{2}>-1$, so we have the standard behaviour of this derivative
at the wormhole throat. On the other hand, we can see from
Eq.~(\ref{embedding}) that the embedding of considered spacetimes in
ordinary three dimensional Euclidean space has a finite size since
it extends from $r_0$ up to a maximum radial value $r_{max}$.
Effectively, the radial coefficient $g_{rr}$ of the
metric~(\ref{metric LM MT}) implies that the denominator of the
fraction is positive for any $r>r_0$, thus the numerator must be
also positive. Then the condition $\beta(\beta-2)\left(\frac{r}{r_0}
\right)^{\frac{2(\beta^2-2\beta-1)}{\beta+1}}\leq1$ must be
required, implying that $r_0\leq r\leq r_{max}$, where
\begin{eqnarray}\label{rmax}
r_{max}=r_0(\beta(\beta-2))^{-\frac{\beta+1}{2(\beta^2-2\beta-1)}}.
\end{eqnarray}
Notice that for any value of the $\beta$-parameter we have that
$\frac{dz(r)}{dr}|_{r_0}=\infty$ and
$\frac{dz(r)}{dr}|_{r_{max}}=0$. It can be shown that for $\beta=-1$
this $r_{max}=r_0$, as well as for $\beta \rightarrow -\infty$ we
have that $r_{max} \rightarrow r_0$. The sphere $r_{max}$ has a
maximum value for $\beta=-4.745695219$ where it takes the value
$r_{max}=1.232835973 \, r_0$. In Fig.~(\ref{Graficoparabeta15}) we
show the behaviour of $r_{max}$ for $\beta\leq -1$ and $r_0=1$.

In order to study the shape and the size of slices $t=const,
\theta=\frac{\pi}{2}$ of the metric~(\ref{metric LM MT}) we shall
consider for the $\beta$-parameter values
$\beta=-2,-3,-4.745695219,-10,-15,-30,-100<-1$. These embeddings are
shown in Fig.~(\ref{embedding-lineas15}). For a full visualization
of the surfaces the diagrams must be rotated about the vertical
$z$-axis. We conclude from these diagrams that for $\beta<-1$ the
metric~(\ref{metric LM MT}) has typical wormhole shapes, i.e.
presence of a global minimum at $r=r_0$, where the throat of the
wormhole is located. The radial extension for all embeddings is
finite as we would expect, and the heigh of the $z$-function
increases with decreasing $\beta$-parameter.

In Fig.~(\ref{beta-615}) we show three-dimensional wormhole
embedding diagrams for the values of the $\beta$-parameter
$-2,-3,-4.745695219,-10$. It becomes clear that all embedding
surfaces flare outward. This can be seen also from the fact that for
the metric~(\ref{metric LM MT}) the inverse of the embedding
function $r(z)$ satisfies
\begin{eqnarray}
d^2r/dz^2= \nonumber
\\ \frac{-r(\beta^2-2\beta-1)^2\left(\frac{r}{r_0}\right)^{2\beta}}{r_0^2(1+\beta)
\left(\left(\frac{r}{r_0}\right)^{\frac{2\beta^2}{1+\beta}}(\beta^2-2
\beta )-\left(\frac{r}{r_0}\right)^{\frac{2(2\beta+1)}{1+\beta}}
\right)^2},
\end{eqnarray}
implying that at the throat we have that
\begin{eqnarray}\label{gargantaderivada}
d^2r/dz^2|_{r_0}=-\frac{1}{r_0 (1+\beta)},
\end{eqnarray}
which is positive for any $\beta<-1$. Thus, the required flare-out
condition for the wormhole throat is satisfied~\cite{Morris:1988cz}.
Note that Eq.~(\ref{gargantaderivada}) implies that for $\beta
\rightarrow -\infty$ we have $d^2r/dz^2|_{r_0} \rightarrow 0$. By
taking into account that we have also that $r_{max} \rightarrow r_0$
if $\beta \rightarrow -\infty$, then we may conclude that the shape
of the wormhole embedding becomes a cylinder of radius $r_0$ for big
negative values of the $\beta$-parameter.

In conclusion, the metric~(\ref{metric LM MT}) describes a wormhole
geometry with isotropic pressure wich extends from $r=r_0$ to
$r=\infty$. Since these wormholes are not asymptotically flat, and
the embedding in ordinary three dimensional Euclidean space extends
from $r_0$ to $r_{max}$, we may match them, as an interior
spacetime, to an exterior vacuum spacetime at the finite junction
surface $r=r_{max}$.

From Eqs.~(\ref{ED}) and~(\ref{PL}) we conclude that at the throat
we have for the energy density that $\rho(r_0)<0$ if $-3<\beta<-1$
and $\rho(r_0) \geq0$ for $\beta\leq-3$, while for the pressure we
obtain $p(r_0)=-1/r_0^2<0$. On the other hand we have that
\begin{eqnarray}
\rho+p=\frac{2\left(\frac{r}{r_0}
\right)^{-\frac{2(\beta^2-2\beta-1)}{1+\beta}}(2\beta+1)(\beta-1)}{(1+\beta)(\beta^2-2\beta-1)r^2}-
\\ \nonumber \frac{2\beta }{(\beta^2-2\beta-1)r^2},
\end{eqnarray}
then at the throat $\rho+p=\frac{2}{(1+\beta)r_0^2}$ is fulfilled.
Thus for $\beta<-1$ we have that always $\rho+p<0$, which allows us
to conclude that the energy conditions are not satisfied at the
wormhole throat.

Since we are interested in studying static wormhole configurations,
for which we must require $\beta<-1$, we conclude that we have
wormholes only exhibiting a solid angle deficit, for which we obtain
that $0<\frac{1}{\beta^2-2 \beta-1}<1/2$. The polar and azimuthal
angles are restricted to $0 \leq \theta \leq \pi$ and $0 \leq \phi <
2 \pi$, respectively, therefore $0 \leq \tilde{\theta} \leq
\pi/\sqrt{2}$ and $0\leq \tilde{\phi} < \sqrt{2} \pi$, where
$\tilde{\theta}=\theta/\sqrt{\beta^2-2 \beta-1}$ and
$\tilde{\phi}=\phi/\sqrt{\beta^2-2 \beta-1}$.



\section{Conclusions}

The study of spherically symmetric traversable wormholes in General
Relativity has been mostly focused in sources with anisotropic
pressures. In this work we present and discuss static spherical
wormhole spacetimes supported by a single perfect fluid, for which
the condition $p_r=p_l$ must be required for radial and lateral
pressures.

We show that it is not possible to sustain a zero-tidal-force
wormhole by a perfect fluid, thus a single fluid threading a
zero-tidal-force wormhole must be necessarily anisotropic. This
implies that if we want to generate spherically symmetric wormholes,
sustained by a single matter source with isotropic pressure, we must
consider spacetimes with $\phi(r) \neq const$.

Also we discuss the possibility of having isotropic fluids with a
linear equations of state. In particular, we show that a wormhole
with a power-law shape function cannot be supported by an ideal
fluid with a linear equation of state. Therefore we consider more
general forms for the shape function and the equation of state of
the isotropic pressure.

In this manner, we generate and discuss the general solution for a
non-asymptotically flat family of static wormholes characterized by
a redshift function given by Eq.~(\ref{ephi}). The obtained wormhole
solutions exhibit always a solid angle deficit and do not satisfy
energy conditions. The embeddings of these spacetimes in ordinary
three dimensional Euclidean space have a finite size since they
extend from $r_0$ up to a maximum radial value $r_{max}$ given by
Eq.~(\ref{rmax}). However, notice that the metric~(\ref{metric LM
MT}), or equivalently the metric~(\ref{metric LM MT varrho}), is
well behaved for $r \geq r_0$, including the sphere $r_{max}$. For
$r \geq r_{max}$ wormhole slices cannot be embedded in an ordinary
Euclidean space. Instead, a space with indefinite metric must be
used. It is interesting to note that one may match such a
non-asymptotically flat wormhole, as an interior spacetime, to an
exterior vacuum spacetime at the finite junction surface
$r=r_{max}$.

\section{Acknowledgements}
This work was supported by Direcci\'on de Investigaci\'on de la
Universidad del B\'\i o-B\'\i o through grants N$^0$ DIUBB 140708
4/R and N$^0$ GI 150407/VC.

\end{document}